\documentclass[12pt,a4paper]{article}

\usepackage{epsfig}
\usepackage{amsmath,amsfonts,amssymb}
\usepackage{t1enc}
\usepackage{cite}

\providecommand{\openone}{\leavevmode\hbox{\small1\kern-3.8pt\normalsize1}}

\newcommand{\ptmiss}{p_T\!\!\!\!\!\!\!\!\not\,\,\,\,\,\,\,}

\begin{document}

\begin{center}
\begin{Large}
{\bf Electroweak scale seesaw and heavy Dirac \\[2mm] neutrino signals at LHC}
\end{Large}

\vspace{0.5cm}
F. del Aguila, J. A. Aguilar--Saavedra  \\[0.2cm] 
{\it Departamento de F\'{\i}sica Te\'orica y del Cosmos and CAFPE, \\
Universidad de Granada, E-18071 Granada, Spain} \\[0.1cm]
\end{center}

\begin{abstract}
Models of type I seesaw can be implemented at the electroweak scale in a natural way provided that the heavy neutrino singlets are quasi-Dirac particles. In such case, their contribution to light neutrino masses has the suppression of a small lepton number violating parameter, so that light neutrino masses can arise naturally even
if the seesaw scale is low and the heavy neutrino mixing is large. We implement the same mechanism with fermionic triplets in type III seesaw,
deriving the interactions of the new quasi-Dirac neutrinos and heavy charged leptons with the SM fermions. We then study the observability of heavy Dirac neutrino singlets (seesaw I) and triplets (seesaw III) at LHC. Contrarily to common wisdom, we find that heavy Dirac neutrino singlets with a mass around 100 GeV are observable at the $5\sigma$ level with a luminosity of 13 fb$^{-1}$. Indeed, in the final state with three charged leptons $\ell^\pm \ell^\pm \ell^\mp$, not previously considered, Dirac neutrino signals can be relatively large and backgrounds are small. In the triplet case, heavy neutrinos can be discovered with a luminosity of 1.5 fb$^{-1}$ for a mass of 300 GeV in the same channel.
\end{abstract}

\section{Introduction}

The well-known seesaw mechanism provides a simple and natural way to give tiny masses to the three neutrinos present in the Standard Model (SM). In its initial realisation (type I seesaw) \cite{Minkowski:1977sc,GellMann:1980vs,Yanagida:1979as,
Mohapatra:1979ia} the SM is enlarged with the addition of three heavy right-handed neutrino singlets $N'_{iR}$ ($i=1,2,3$). These fields have Yukawa interactions with the SM left-handed lepton doublets $L'_{iL}$ as well as a Majorana mass term, 
\begin{equation}
\mathcal{L}^\text{I}_\text{mass} = -Y_{ij} \, \bar L'_{iL} N'_{jR} \, \tilde \phi
-\frac{1}{2} M_{ij} \bar N'_{iL} N'_{jR}
 +\text{H.c.}\,, 
\end{equation}
where $\phi$ is the SM Higgs doublet, $N'_{iL} \equiv N_{iR}^{'c} = (N'_{iR})^c$ and $\tilde \phi=i\tau_2 \phi^*$, with $\tau_i$ the Pauli matrices. After electroweak symmetry breaking, these terms generate a mass matrix for neutrino fields and, in particular, an effective light neutrino mass
matrix
\begin{equation}
M_\nu = -\frac{v^2}{2} Y M^{-1} Y^T \,,
\label{ec:mnu}
\end{equation}
where $v=246$ GeV is the vacuum expectation value of the SM Higgs.
For Yukawa couplings $Y_{ij}$ of order one without any particular symmetry, light neutrino masses $m_\nu \sim 0.1$ eV can be obtained if the heavy neutrino masses
(which are approximately equal to the eigenvalues of $M$) are of the order $m_N \sim 10^{14}$ GeV, close to the unification scale. However, the drawback of this picture is that these new eigenstates are too heavy to be observable, and this ``minimal'' type I seesaw mechanism cannot be directly tested. Many attempts have been made in the literature to lower the heavy scale while successfully generating light neutrino masses compatible with experimental data. One possibility is to have very small Yukawa couplings. Indeed, in the SM only the top quark has a Yukawa coupling of order unity, while for the charged leptons they range from $2.9 \times 10^{-6}$ for the electron to $10^{-2}$ for the tau. But heavy 
neutrinos are still unobservable in this case because their mixing with the SM leptons,
\begin{equation}
V_{lN} \simeq \frac{Y_{lN} v}{\sqrt 2 m_N} \,,
\end{equation}
is very small and they cannot be directly produced at an observable rate. For example, for heavy neutrino masses $m_N \sim v$ the requirement $m_\nu \lesssim 0.1$ eV implies $V_{lN} \lesssim 10^{-6}$, giving a suppression factor of $|V_{lN}|^2 \sim 10^{-12}$ for heavy neutrino production cross sections. A more interesting possibility is to keep sizeable Yukawa couplings but introduce some symmetry which suppresses the seesaw-type contribution in Eq.~(\ref{ec:mnu}). In the examples known in the literature~\cite{Buchmuller:1991tu,Ingelman:1993ve,Tommasini:1995ii,
Pilaftsis:2004xx,Pilaftsis:2005rv,Kersten:2007vk,Abada:2007ux},
this symmetry is lepton number.
In this way, two nearly-degenerate heavy Majorana neutrinos with masses $m_N \pm \mu/2$ and opposite CP parities form a quasi-Dirac neutrino, whose contribution to light neutrino masses has the further suppression of the small ratio $\mu/m_N$ (see~\cite{delAguila:2008ks} and references there in). Then, this quasi-Dirac neutrino can have a relatively large mixing with the SM leptons, only constrained by electroweak precision observables~\cite{delAguila:2008pw}.

The same mechanism can be implemented with fermionic triplets in type III seesaw. In its usual realisation \cite{Foot:1988aq,Ma:1998dn}, type III seesaw introduces three fermion triplets $\Sigma_i$ ($i=1,2,3$) with zero hypercharge,
each of them consisting of one Majorana neutrino $N_i$ and two Weyl spinors which can form a Dirac charged lepton $E_i$. Fermion triplets couple to the SM lepton doublets through Yukawa couplings and have Majorana mass terms as well. Similarly to type I seesaw, light neutrino masses are of order $m_\nu \sim Y^2 v^2 / (2 m_N)$ and the mixing of the heavy states is $V \sim Yv / (\sqrt 2 m_N)$ (see next section). If the new particles are near the electroweak scale, light neutrino masses can be kept within experimental limits either with $Y$ very small\footnote{Note that a sizeable mixing $V$ is not necessary to observe the new heavy states, because they can be produced in pairs by unsuppressed gauge interactions and decay within the detector for $Y \gtrsim 10^{-6}$ \cite{Franceschini:2008pz}.} or by requiring some symmetry. 
In this Letter we work out in detail the second possibility. In full analogy with type I seesaw, if the heavy neutrinos are quasi-Dirac particles the light neutrino masses get a further suppression $\mu/m_N$, so that Yukawa couplings of order unity are 
possible with $m_N$ at the electroweak scale. In this ``Dirac'' variant of type III seesaw each heavy quasi-Dirac neutrino $N$ is associated with two charged leptons $E_1^-$ and $E_2^+$, the three of them with lepton number equal to one and nearly degenerate in mass,
$m_N \simeq m_{E_1^-} \simeq m_{E_2^+}$. Lepton number is conserved except for the light neutrino Majorana masses. 
We obtain the interactions between the new states $N$, $E_1^-$, $E_2^+$ and the SM leptons, finding substantial differences with the Majorana case, which affect their decay channels and the possible signals at colliders. It must be remarked that the seesaw mechanism with heavy quasi-Dirac neutrinos is naturally realised in models with new physics at the electroweak scale, like in little Higgs models with simple groups~\cite{delAguila:2005yi} or in models with extra dimensions and matter in the bulk implying infinite Kaluza-Klein towers of Dirac fermions (see for example Refs.~\cite{Grossman:1999ra,Csaki:2003sh}).

After describing in detail this variant of type III seesaw, we study the observability of heavy Dirac neutrino singlets and triplets at the Large Hadron Collider (LHC). We follow the procedure previously used in Ref.~\cite{delAguila:2008cj}, and systematically study all the observable multi-lepton signals (up to six leptons)
including all signal contributions from the various production processes and decay channels. By comparing with the ``minimal'' seesaw I and III scenarios  with heavy Majorana neutrinos studied in Ref.~\cite{delAguila:2008cj} we show that the discrimination between the two possibilities is very easy, provided that a positive signal is observed. In particular we also find that, in contrast with common wisdom, heavy Dirac neutrino singlets could be observed at LHC in the trilepton final state $\ell^\pm \ell^\pm \ell^\mp$. For example, a heavy neutrino with $m_N = 100$ GeV coupling to the muon with $V_{\mu N}^2 = 0.0032$ can be seen with $5\sigma$ significance with only 13 fb$^{-1}$ of luminosity. For a 300 GeV Dirac fermion triplet the discovery is possible with only 1.5 fb$^{-1}$.

\section{Heavy Dirac neutrinos in type III seesaw}

Here we will follow the notation in Ref.~\cite{delAguila:2008cj} and quote some results obtained there without proof. Fermion triplets $\Sigma_i$, $i=1,\dots,n$ (we leave their number arbitrary) are composed of
three Weyl spinors of zero hypercharge, and we choose them to be right-handed under Lorentz transformations. They couple to the SM lepton doublets 
$L'_{iL}=(\nu'_{iL} \; l'_{iL})^T$ (we use primes for weak interaction eigenstates when necessary to distinguish them from mass eigenstates) and have a Majorana mass term,
\begin{equation}
\mathcal{L}^\text{III}_\text{mass} = -Y_{ij} \, \bar L'_{iL} (\vec \Sigma_j \cdot \vec \tau)
\, \tilde \phi
-\frac{1}{2} \,  M_{ij} \, \overline{\vec \Sigma_i^c} \cdot \vec \Sigma_j
+\text{H.c.} \,,
\label{ec:LmassIII}
\end{equation}
where
$\vec \Sigma_j = (\Sigma_j^1 ,\, \Sigma_j^2 ,\, \Sigma_j^3)$ are the triplets written in Cartesian components. Notice that all the members 
$\Sigma_j^1$, $\Sigma_j^2$, $\Sigma_j^3$ of the triplet $\Sigma_j$ have the same mass, and the matrix $M$ (which is symmetric) can be assumed diagonal in full generality. For each triplet $\Sigma_j$, the charge eigenstates are related to the Cartesian components by
\begin{equation}
\Sigma_j^+ = \frac{1}{\sqrt 2} (\Sigma_j^1-i \Sigma_j^2) \,, \quad
\Sigma_j^0 = \Sigma_j^3 \,, \quad
\Sigma_j^- = \frac{1}{\sqrt 2} (\Sigma_j^1+i \Sigma_j^2) \,,
\label{ec:T:chdef}
\end{equation}
and from them we can build one charged Dirac and one neutral Majorana fermion,
\begin{equation}
E'_j = \Sigma_j^- + \Sigma_j^{+c} \,,\quad N'_j = \Sigma_j^0 + \Sigma_j^{0c} \,.
\label{ec:T:EN}
\end{equation}
With our choice of right-handed chirality for $\Sigma_j$, we have
\begin{equation}
E'_{jL} = \Sigma_j^{+c} \,,\quad E'_{jR} = \Sigma_j^- \,,\quad
N'_{jL} = \Sigma_j^{0c} \,,\quad N'_{jR} = \Sigma_j^{0} \,.
\end{equation}
After spontaneous symmetry breaking, the terms in Eq.~(\ref{ec:LmassIII}) yield the neutrino and charged lepton mass matrices
\begin{eqnarray}
\mathcal{L}_{\nu,\mathrm{mass}} & = & - \frac{1}{2} \,
\left(\bar \nu'_L \; \bar N'_L \right)
\left( \! \begin{array}{cc}
0 & \frac{v}{\sqrt 2} Y \\ \frac{v}{\sqrt 2} Y^T & M
\end{array} \! \right) \,
\left( \!\! \begin{array}{c} \nu'_R \\ N'_R \end{array} \!\! \right)
\; + \mathrm{H.c.} \,, \nonumber \\
\mathcal{L}_{l,\mathrm{mass}} & = & -
\left(\bar l'_L \; \bar E'_L \right)
\left( \! \begin{array}{cc}
\frac{v}{\sqrt 2} Y^l & v Y \\ 0 & M
\end{array} \! \right) \,
\left( \!\! \begin{array}{c} l'_R \\ E'_R \end{array} \!\! \right)
\; + \mathrm{H.c.} \,,
\label{ec:m2x2}
\end{eqnarray}
where we have defined $\nu'_{iR} \equiv \nu^{'c}_{iL}$, $N'_{iL} \equiv N^{'c}_{iR}$.
Note that we have not specified the number of triplets, so $Y$ is a matrix of dimension $3\times n$ and $M$ is $n \times n$.
The mass eigenstates are obtained from the diagonalisation of these mass matrices. Their interactions with the SM leptons (with $l=e,\mu,\tau$) are given by~\cite{delAguila:2008cj}
\begin{align}
\mathcal{L}_W & = - g \left( \bar E_i \gamma^\mu N_i \, W_\mu^-
   + \bar N_i \gamma^\mu E_i \, W_\mu^+ \right) \notag \\
&  - \frac{g}{\sqrt 2} \left( V_{l N_i} \, \bar l \gamma^\mu  P_L N_i \; W_\mu^- 
   + V_{l N_i}^* \, \bar N_i \gamma^\mu  P_L l \; W_\mu^+ \right) \notag \\
&  - g \left( V_{lN_i} \, \bar E_i \gamma^\mu  P_R \nu_l \; W_\mu^- 
    + V_{lN_i}^* \, \bar \nu_l \gamma^\mu  P_R E_i \; W_\mu^+ \right) \,, \notag \\[1mm]
\mathcal{L}_Z & = g c_W \, \bar E_i \gamma^\mu E_i \, Z_\mu \notag \\
&  + \frac{g}{2c_W} \left( V_{lN_i} \, \bar \nu_l \gamma^\mu P_L N_i
+ V_{lN_i}^* \, \bar N_i \gamma^\mu P_L \nu_l \right) Z_\mu \notag \\
&  + \frac{g}{\sqrt 2 c_W} \left( V_{lN_i} \, \bar l \gamma^\mu P_L E_i
+ V_{lN_i}^* \, \bar E_i \gamma^\mu P_L l \right) Z_\mu \,, \notag \\[1mm]
\mathcal{L}_\gamma & = e \, \bar E_i \gamma^\mu E_i \, A_\mu \,, \notag \\[1mm]
\mathcal{L}_H & =  \frac{g \, m_{N_i}}{2 M_W} \left( V_{lN_i} \, \bar \nu_l  P_R N_i
+ V_{lN_i}^* \, \bar N_i  P_L \nu_l \right) H \nonumber \\
& +\frac{g \, m_{E_i}}{\sqrt 2 M_W} \left( V_{lN_i} \, \bar l  P_R E_i
+ V_{lN_i}^* \, \bar E_i  P_L l \right) H \,,
\label{ec:TintM}
\end{align}
where the mixing is
\begin{equation}
V_{lN_i} \simeq - \frac{Y_{lN_i} v}{\sqrt 2 m_{N_i}} \,.
\end{equation}

In the general Majorana case, the mass eigenstates are $E_i$, $N_i$. This still holds when two Majorana neutrinos form a Dirac fermion, but in this case 
the two degenerate Majorana states interfere, and further redefinitions are convenient to simplify the analysis. Then, let us assume that we have two Majorana neutrinos $N_1$ and $N_2$ degenerate in mass and with opposite CP parities (the small mass difference in the quasi-Dirac case is crucial in order to generate light neutrino masses but irrelevant for heavy lepton interactions). Moreover, their Yukawa couplings satisfy $Y_{lN_2} = i Y_{lN_1}$, so that
\begin{equation}
V_{lN_2} = i V_{lN_1} \,.
\label{ec:Vrel}
\end{equation}
The calculation of quantities such as cross sections for heavy neutrino production with subsequent decay can be done in terms of two interfering Majorana neutrinos $N_1$ and $N_2$ (each with two degrees of freedom). Alternatively, it can be done in terms of a Dirac neutrino with components
\begin{equation}
N_L \equiv \frac{1}{\sqrt 2} (N_{1L} + i N_{2L})  \,, \quad
N_R \equiv \frac{1}{\sqrt 2} (N_{1R} + i N_{2R})  \,,
\label{ec:Ndirdef}
\end{equation}
which has four degrees of freedom. The charged sector can also be described in terms of the two interfering degenerate states $E_1$, $E_2$ or in terms of two fermions
$E_1^-$, $E_2^+$ (the latter is positively charged), which are defined as
\begin{align}
E_{1L}^- & \equiv \frac{1}{\sqrt 2} (E_{1L} + i E_{2L}) \,,\quad
E_{1R}^- \equiv \frac{1}{\sqrt 2} (E_{1R} + i E_{2R}) \,, \notag \\
E_{2L}^+ & \equiv \frac{1}{\sqrt 2} (E_{1R}^c + i E_{2R}^c) \,,\quad
E_{2R}^+ \equiv \frac{1}{\sqrt 2} (E_{1L}^c + i E_{2L}^c) \,.
\label{ec:Edirdef}
\end{align}
These states have diagonal couplings to the Dirac neutrino $N$ and do not interfere in the processes analysed in this paper, so the description turns out to be much simpler.
The interactions of the Dirac mass eigenstate fermions $N$, $E_1^-$ and $E_2^+$ are obtained from Eqs.~(\ref{ec:TintM}) for $i=1,2$ using Eqs.~(\ref{ec:Vrel})-(\ref{ec:Edirdef}),
\begin{align}
\mathcal{L}_W & = - g \left( \, \bar E_1^- \gamma^\mu N -\bar N \gamma^\mu E_2^+
\right) W_\mu^-
  -g \left( \bar N \gamma^\mu E_1^- -\bar E_2^+ \gamma^\mu N \right)  W_\mu^+  \notag \\
&  - \frac{g}{\sqrt 2} \left( V_{l N} \, \bar l \gamma^\mu  P_L N \; W_\mu^- 
   + V_{l N}^{*} \, \bar N \gamma^\mu  P_L l \; W_\mu^+ \right) \notag \\
&  + g \left( V_{lN} \, \bar \nu_l \gamma^\mu  P_L E_2^+ \; W_\mu^- 
    + V_{lN}^{*} \, \bar E_2^+ \gamma^\mu  P_L \nu_l \; W_\mu^+ \right) \,,
      \notag \\[1mm]
%%%%
\mathcal{L}_Z & = g c_W \left( \, \bar E_1^- \gamma^\mu E_1^-
-\bar E_2^+ \gamma^\mu E_2^+ \right) Z_\mu \notag \\
&  + \frac{g}{2c_W} \left( V_{lN} \, \bar \nu_l \gamma^\mu P_L N
+ V_{lN}^{*} \, \bar N \gamma^\mu P_L \nu_l \right) Z_\mu \notag \\
&  + \frac{g}{\sqrt 2 c_W} \left( V_{lN} \, \bar l \gamma^\mu P_L E_1^-
+ V_{lN}^{*} \, \bar E_1^- \gamma^\mu P_L l \right) Z_\mu \,, \displaybreak \notag \\[1mm]
%%%%
\mathcal{L}_\gamma & = e \left( \, \bar E_1^- \gamma^\mu E_1^-
-\bar E_2^+ \gamma^\mu E_2^+ \right) A_\mu \,, \notag \\[1mm]
%%%%
\mathcal{L}_H & =  \frac{g \, m_N}{2 M_W} \left( V_{lN} \, \bar \nu_l  P_R N
+ V_{lN}^{*} \, \bar N  P_L \nu_l \right) H \nonumber \\
& +\frac{g \, m_{E_1}}{\sqrt 2 M_W} \left( V_{lN} \, \bar l  P_R E_1^-
+ V_{lN}^{*} \, \bar E_1^-  P_L l \right) H \,,
\label{ec:TintD}
\end{align}
with $m_N = m_{E_1^-} = m_{E_2^+}$ and $V_{lN} = \sqrt 2 V_{lN_1}$.
The Feynman rules for these interactions are collected in the Appendix.
The main differences with the Majorana case are:
\begin{itemize}
\item[(i)] the gauge couplings with two heavy fermions have opposite sign for $E_2^+$ and $E_1^-$;
\item[(ii)] light neutrinos only couple to $E_2^+$ , while light charged leptons only couple to $E_1^-$. 
\end{itemize}
On the other hand, the heavy Dirac neutrino couples to light charged and neutral leptons in the same way as a heavy Majorana neutrino, except that in the latter case the neutral couplings can be rewritten using $N=N^c$, $\nu=\nu^c$.
These equations also show clearly that, if lepton number $\mathrm{L}=1$ is assigned to $N$, $E_1^-$ and $E_2^+$, then it is conserved except for light neutrino masses. When the two Majorana states are not degenerate but have masses $m_N \pm \mu/2$ and $N$ is quasi-Dirac, lepton number violating (LNV) effects are of order $\mu/m_N$.

There are several implementations of this mechanism in the heavy lepton mass matrices (see for example Refs.~\cite{Buchmuller:1991tu,Ingelman:1993ve,Tommasini:1995ii,Pilaftsis:2004xx,
Pilaftsis:2005rv,Kersten:2007vk,Abada:2007ux}).
A particularly simple model can be built by assigning lepton number to the heavy triplets and requiring approximate lepton number conservation.
Let us assume that we have two triplets $\Sigma_1$ and $\Sigma_2$, to which we assign lepton numbers equal to $1$ and $-1$, respectively. Then, for one lepton generation the fields are 
\begin{equation}
l'_L \,,\; l'_R \,,\; E'_{1R}  \,,\; E'_{2L} \,,\; \nu'_L \,,\; N'_{1R} \,,\;  \quad (\mathrm{L}=1) \;; \quad \quad \quad
N'_{2R} \,,\; E'_{1L} \,,\; E'_{2R} \quad (\mathrm{L}=-1) \,,
\end{equation}
whereas the charge conjugate fields $\nu'_R \equiv \nu_L^{'c}$, $N'_{2L} \equiv N_{2R}^{'c}$, $N'_{1L} \equiv N_{1R}^{'c}$ have opposite lepton number.
If we impose lepton number conservation, for one lepton generation the neutrino mass term is \cite{Mohapatra:1986aw}
\begin{eqnarray}
\mathcal{L}_{\nu,\mathrm{mass}} & = & - \frac{1}{2} \,
\left(\bar \nu'_L \; \bar N'_{1L} \; \bar N'_{2L} \right)
\left( \! \begin{array}{ccc}
0 & v Y & 0 \\ v Y & 0 & M \\ 0 & M & 0
\end{array} \! \right) \,
\left( \!\! \begin{array}{c} \nu'_R \\ N'_{1R} \\ N'_{2R} \end{array} \!\! \right)
\; + \mathrm{H.c.} 
\label{ec:Mndir}
\end{eqnarray}
The $2 \times 2$ submatrix in the heavy sector corresponds to the mass term of a heavy Dirac neutrino $N = N'_{2L} + N'_{1R}$ with mass $m_N = M$,
\begin{equation}
-\frac{M}{2}  \left( \bar N'_{1L} N'_{2R} + \bar N'_{2L} N'_{1R} \right)
= -\frac{M}{2} \left( \overline{N_{1R}^{'c}} N_{2L}^{'c} + N'_{2L} N'_{1R} \right)
= - M \bar N'_{2L} N'_{1R} \,.
\end{equation}
This definition is consistent with Eqs.~(\ref{ec:Ndirdef}). A tiny light neutrino Majorana mass 
\begin{equation}
m_\nu = \frac{\mu v^2 Y^2}{2 M^2}
\end{equation}
can be introduced with a LNV entry $\mu \ll M$ in the $(3,3)$ position of this matrix, in which case the heavy neutrino is quasi-Dirac, being the masses of the two heavy Majorana fields $M \pm \mu/2$. For three lepton generations and six fermion triplets (in order to yield three heavy Dirac neutrinos) the structure of the neutrino mass matrix is the same but $Y$ and $M$ are $3 \times 3$ matrices, and they have to be replated by $Y^T$ and $M^T$ in the $(2,1)$ and $(3,2)$ positions of the matrix in Eq.~(\ref{ec:Mndir}), respectively.

In the charged sector the mass term is, for one lepton generation,
\begin{eqnarray}
\mathcal{L}_{l,\mathrm{mass}} & = & -
\left(\bar l'_L \; \bar E'_{1L} \; \bar E'_{2L} \right)
\left( \! \begin{array}{ccc}
\frac{v}{\sqrt 2} Y^l & \sqrt 2 v Y & 0 \\ 0 & 0 & M \\ 0 & M & 0 \end{array} \! \right) \,
\left( \!\! \begin{array}{c} l'_R \\ E'_{1R} \\ E'_{2R} \end{array} \!\! \right)
\; + \mathrm{H.c.} 
\end{eqnarray}
By inspection of the mass matrix one observes that the $2 \times 2$ submatrix in the heavy sector corresponds to the diagonal mass term of two Dirac fermions $E_1^- = E'_{2L} + E'_{1R}$, $E_2^+ = E_{1L}^{'c} + E_{2R}^{'c}$, both with $\text{L}=1$
and $m_{E_1^-} = m_{E_2^+} = M$.
This is consistent with the definition of fields in Eqs.~(\ref{ec:Edirdef}).

We finally point out that for collider phenomenology (where the energy scale is much larger than light neutrino masses, which can then be neglected), Dirac and quasi-Dirac neutrinos are equivalent, thus in our simulations we assume that heavy neutrinos are Dirac particles.

\section{Heavy Dirac neutrino signals at LHC}

\begin{figure}[ht]
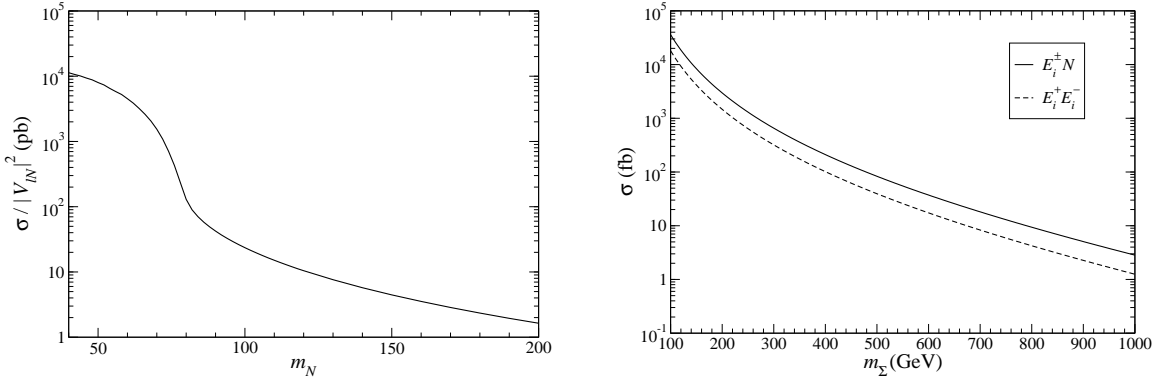

\begin{center}
\begin{tabular}{ccc}
\epsfig{file=Figs/cross-singlet.eps,height=5cm,clip=} & \quad &
\epsfig{file=Figs/cross-fermion.eps,height=5cm,clip=}
\end{tabular}
\caption{Cross sections for heavy Dirac neutrino singlet (left) and fermion triplet production (right). In the latter case the production of both charged leptons $E_1^-$ and $E_2^+$ is summed and the triplet masses are taken equal, $m_\Sigma \equiv m_N = m_{E_i^\pm}$.}
\label{fig:cross}
\end{center}
\end{figure}

The production processes of heavy Dirac neutrino singlets and triplets are the same as for their Majorana counterparts.\footnote{The transition between the Dirac and Majorana cases can be made smooth by varying continuously the couplings and masses in Eq.~(\ref{ec:m2x2}), as it happens, for instance, for gluinos and neutralinos in supersymmetry~\cite{Choi:2008pi}.}
Dirac neutrino singlets can be produced at hadron colliders in the process
\begin{equation}
q \bar q' \to W^{*} \to l^\pm N \,.
\label{ec:pr-sin}
\end{equation}
They can also be produced in $q \bar q \to Z^* \to \nu N$ but the resulting signal has at most one charged lepton, and is swamped by the background.
The charged and neutral members of a Dirac triplet can be produced in
\begin{align}
& q \bar q \to Z^* \,/\, \gamma^* \to E_i^+ E_i^- \,, \notag \\
& q \bar q' \to W^* \to E_i^\pm N \,,
\label{ec:pr-tri}
\end{align}
where $E_i^\pm$ refers here to the $\mathrm{L}=1$ fermions $E_1^-$ and $E_2^+$ defined in the previous section. The cross sections of these processes are shown in Fig.~\ref{fig:cross}.
The decays are different in the Dirac and Majorana cases, however.
Heavy Dirac neutrinos decay in the channels $N \to W^+ l^-$, $N \to Z \nu_l$ and $N \to H \nu_l$, but not in the LNV mode $N \to W^- \ell^+$, which leads to like-sign dilepton signals in $l^\pm N$ production when $N$ is a Majorana fermion (for a complete study see Ref.~\cite{delAguila:2007em}). The interactions of a heavy Dirac neutrino singlet can be found in Ref.~\cite{delAguila:2005pf}, and the relevant Feynman rules are collected in the Appendix.
The partial widths for $N$ decays, both in the singlet and triplet cases, are
\begin{eqnarray}
\Gamma(N \to l^- W^+) & = & \frac{g^2}{64 \pi} |V_{l N}|^2
\frac{m_N^3}{M_W^2} \left( 1- \frac{M_W^2}{m_N^2} \right) 
\left( 1 + \frac{M_W^2}{m_N^2} - 2 \frac{M_W^4}{m_N^4} \right) \,, \nonumber
\\[0.1cm]
\Gamma(N \to \nu_l Z) & = &  \frac{g^2}{128 \pi c_W^2} |V_{lN}|^2
\frac{m_N^3}{M_Z^2} \left( 1- \frac{M_Z^2}{m_N^2} \right) 
\left( 1 + \frac{M_Z^2}{m_N^2} - 2 \frac{M_Z^4}{m_N^4} \right) \,, \nonumber
\\[0.2cm]
\Gamma(N \to \nu_l H) & = &  \frac{g^2}{128 \pi} |V_{lN}|^2
\frac{m_N^3}{M_W^2} \left( 1- \frac{M_H^2}{m_N^2} \right)^2 \,.
\label{ec:Nwidths}
\end{eqnarray}
For heavy charged leptons the decays are different in the Dirac case too.
Since $E_2^+$ does not couple to charged leptons, it always decays in the $E_2^+ \to \nu_l W^+$ channel. The opposite happens for $E_1^-$, which does not decay in this mode but in $E_1^- \to l^- \, Z/H$. The corresponding partial widths are
\begin{align}
\Gamma(E_2^+ \to \nu_l W^+ ) & = \frac{g^2}{32 \pi} |V_{lN}|^2
\frac{m_E^3}{M_W^2} \left( 1- \frac{M_W^2}{m_E^2} \right) 
\left( 1 + \frac{M_W^2}{m_E^2} - 2 \frac{M_W^4}{m_E^4} \right) \,, \nonumber
\\[0.1cm]
\Gamma(E_1^- \to l^- Z) & =  \frac{g^2}{64 \pi c_W^2} |V_{lN}|^2
\frac{m_E^3}{M_Z^2} \left( 1- \frac{M_Z^2}{m_E^2} \right) 
\left( 1 + \frac{M_Z^2}{m_E^2} - 2 \frac{M_Z^4}{m_E^4} \right) \,, \nonumber
 \\[0.2cm]
\Gamma(E_1^- \to l^- H) & =  \frac{g^2}{64 \pi} |V_{lN}|^2
\frac{m_E^3}{M_W^2} \left( 1- \frac{M_H^2}{m_E^2} \right)^2 \,.
\label{ec:Ewidths}
\end{align}

The simulation of the seesaw I and III signals with Majorana neutrinos and the SM backgrounds has been previously performed in Ref.~\cite{delAguila:2008cj}. Here we generate the signals corresponding to Dirac neutrinos and compare with the former ones, taken from that study.
In our simulations,
the heavy Dirac neutrino singlet signal in Eq.~(\ref{ec:pr-sin}) is calculated with the {\tt Alpgen} extension in Ref.~\cite{delAguila:2007em}. The triplet signals in Eqs.~(\ref{ec:pr-tri}) are calculated implementing the processes for Dirac neutrinos in Eqs.~(\ref{ec:pr-tri}) in the generator {\tt Triada}~\cite{delAguila:2008cj}.
The matrix elements for the $2 \to 6$ processes are calculated taking finite width and spin effects into account.
All the decay channels of the heavy leptons and the subsequent $W/Z/H$ boson decays
are included, which is a cumbersome task involving 144, 25 and 578 final states for
$E_1^+ E_1^-$, $E_2^+ E_2^-$ and $E_i^\pm N$ production, respectively. 
SM backgrounds include $t \bar t nj$ (where $nj$ stands for $n$ additional jets at partonic level), single top, $W/Z \, nj$, $W/Z b \bar b \, nj$, diboson and triboson production plus jets, and other less important processes. They are all generated with {\tt Alpgen}~\cite{Mangano:2002ea}. Signal and background events are passed through the parton shower Monte Carlo {\tt Pythia}~\cite{Sjostrand:2006za} and a fast simulation of the ATLAS detector~\cite{atlfast}.
Further details about the signal and background generation can be found in Ref.~\cite{delAguila:2008cj}.
As it has been done in that work, here we classify signals by lepton multiplicity and compare the possible final states to find the differences between the Dirac and Majorana cases.

It must be pointed out that, since the Dirac and Majorana neutrino amplitudes are different (although the interactions are formally the same, in the latter case $N=N^c$), one expects to find different angular distributions, for example the forward-backward asymmetry in the $W$ rest frame $A_\text{FB}$ \cite{delAguila:2005pf}. Nevertheless, measuring this asymmetry is very difficult. Its value is proportional to $M_W^2 / m_N^2$, and for a 300 GeV Dirac neutrino it is already very small, $A_\text{FB} = 0.094$, and measuring it requires high statistics. Moreover, its measurement requires either to distinguish between the two quarks produced in the hadronic $W$ decay, or, in leptonic $W$ decays, to distinguish between the charged lepton produced in $W \to \ell \nu$ and the other leptons present in the event. For charged heavy leptons $E_i^\pm$ the decay matrix elements are the same in the Dirac and Majorana cases, so that asymmetries are equal. This happens even in the charged current decay of $E_2^+$, where the chirality of the coupling involved is different in the Dirac and Majorana cases but the charge of the heavy lepton too, with the net result that the amplitudes are the same.
As we will find, the discrimination between Dirac and Majorana neutrinos in terms of charged lepton multiplicities is much more interesting.

\subsection{Neutrino singlets}
\label{sec:3.1}

For Dirac and Majorana neutrino singlets we examine final states with (a) three leptons $\ell^\pm \ell^\pm \ell^\mp X$ and (b) two like-sign leptons $\ell^\pm \ell^\pm X$, where $\ell=e,\mu$ and $X$ denotes possible additional jets. We take the same heavy neutrino parameters used in Ref.~\cite{delAguila:2008cj} for Majorana neutrinos, that is, $m_N = 100$ GeV and two scenarios: $N$ coupling only to the electron with $V_{eN}^2=0.0030$ (S1) or, alternatively, only to the muon with $V_{\mu N}^2 = 0.0032$ (S2). We also use the same pre-selection and selection criteria and reconstruction cuts, summarised below for completeness. Further details, in particular regarding the mass reconstruction, can be found in Ref.~\cite{delAguila:2008cj}.
\begin{itemize}
\item[(a)] {\em Three leptons $\ell^\pm \ell^\pm \ell^\mp$}. Pre-selection: three leptons with total charge $\pm 1$, the two like-sign ones having transverse momentum $p_T > 30$ GeV. Two event samples are considered, one with two or more electrons ($2e$) and the other one with two or more muons ($2\mu$). Selection and reconstruction: the opposite pairs cannot have an invariant mass closer to $M_Z$ than 10 GeV; no $b$ jets can be present; the transverse angle between the like-sign pair $\phi_T^{\ell \ell}$ must be larger than $\pi/2$; missing energy $\ptmiss < 100$ GeV; reconstructed mass of the heavy neutrino $m_N^\text{rec} < 125$ GeV.
\item[(b)] {\em Two like-sign leptons $\ell^\pm \ell^\pm$}. Pre-selection: two like-sign leptons with $p_T > 30$ GeV. Two different event samples are considered: with two electrons and with two muons. Selection and reconstruction: at least two jets with $p_T > 20$ GeV, with no $b$-tagged jets; $\ptmiss < 30$ GeV; $\phi_T^{\ell\ell} < \pi/2$; $m_N^\text{rec} < 180$ GeV.
\end{itemize}
The number of events for 30 fb$^{-1}$ with these criteria is collected in Table~\ref{tab:comp-S} for the heavy neutrino signal in scenarios S1 and S2, in case that the heavy neutrino has Majorana or Dirac character. The SM background is also included.
\begin{table}[b]
\begin{center}
\begin{tabular}{ccccc}
           & $\ell^\pm \ell^\pm \ell^\mp$ ($2e$) & $\ell^\pm \ell^\pm \ell^\mp$ ($2\mu$) & $\ell^\pm \ell^\pm$ ($2e$) & $\ell^\pm \ell^\pm$ ($2\mu$) \\
\hline
$N$ (S1,M) &  28.6  & 0 & 11.3 & 0 \\
$N$ (S1,D) &  44.8  & 0 & 0.4  & 0  \\
$N$ (S2,M) &  0  & 29.6 & 0 & 13.4 \\
$N$ (S2,D) &  0  & 45.8 & 0  & 0.5  \\
SM Bkg  &  116.4 & 45.6 & 36.1 & 20.2
\end{tabular}
\caption{Number of events with 30 fb$^{-1}$ for the Majorana (M) and Dirac (D) neutrino singlet signals in scenarios S1 and S2, and SM background in different final states.}
\label{tab:comp-S}
\end{center}
\end{table}
It is very interesting to observe that heavy Dirac neutrinos could be observed at LHC in the trilepton final state. This fact contradicts previous common wisdom that heavy Dirac neutrino signals are overwhelmed by the SM background (which is actually huge for the final state with two opposite sign leptons $\ell^+ \ell^-$). For a Dirac neutrino with $m_N = 100$ GeV the trilepton signal is actually larger than for a 
Majorana neutrino. This is easily understood from kinematics. The two processes giving trilepton signals are
\begin{align}
& q \bar q' \to \ell^+ N \to \ell^+ \ell^- W^+ \to \ell^+ \ell^- \ell^+ \nu \quad \text{(LNC)} \,, \notag \\
& q \bar q' \to \ell^+ N \to \ell^+ \ell^+ W^- \to \ell^+ \ell^+ \ell^- \bar \nu \quad \text{(LNV)} \,,
\label{ec:Nsinch}
\end{align}
plus the charge conjugate.
In the LNV process the lepton from $N$ decay (which has the same charge as the one produced in association with $N$) is very soft due to the small $m_N-M_W$ mass difference, so this signal is very suppressed by the cut $p_T > 30$ GeV on like-sign leptons. (This cut is necessary to suppress the $t \bar t nj$ background, where a softer like-sign lepton results from the decay of a $b$ quark.) Actually, for a Majorana neutrino most of the events in this final state come from the lepton number conserving (LNC) process, where the like-sign lepton results from $W$ decay. For a Dirac neutrino the LNV decay is absent and the LNC decay has a branching ratio twice as large, so that the trilepton signal is enhanced. Assuming a 20\% background uncertainty, the Dirac heavy neutrino signal could be observed with $5\sigma$ significance with a luminosity of 13 fb$^{-1}$ in scenario S2, while for scenario S1 the discovery significance cannot be achieved unless the background is known with a better precision.

Conversely, like-sign dilepton signals, which have similar significance as trilepton ones for a Majorana neutrino, are practically absent for a Dirac neutrino
(like-sign dileptons can still be produced with a much lower rate from the LNC decay in Eq.~(\ref{ec:Nsinch}) when $\ell^-$ is missed by the detector).
Then, a Dirac neutrino might be identified with an event excess in the trilepton channel without any like-sign dilepton signal, while a Majorana neutrino would give signals in both channels with similar significance~\cite{delAguila:2008cj}.

\subsection{Fermion triplets}

Fermion triplet production gives signals with up to six leptons. The differences with respect to the Majorana case are originated by the absence of the LNV decay $N \to \ell^+ W^-$ and of the mixed decays of an $E^+ E^-$ pair giving a $W^\pm$ and a $Z/H$ boson. The cross section is two times larger in the Dirac case, because $E_1^-$ and $E_2^+$ are both produced. We take a triplet mass of 300 GeV, and assume that it only couples to the electron (for a coupling to the muon or to both the results are very similar, as argued in Ref.~\cite{delAguila:2008cj}).
The final states examined, with a summary of pre-selection and selection criteria and reconstruction cuts, if any, are:
\begin{itemize}
\item[(a)] {\em Six charged leptons}. Pre-selection and selection: six charged leptons, two of them with $p_T > 30$ GeV and the rest with $p_T > 15$ GeV (for electrons) and $p_T > 10$ GeV (for muons).
\item[(b)] {\em Five charged leptons}. The same as above but with five leptons instead of six.
\item[(c)] {\em Four leptons $\ell^\pm \ell^\pm \ell^\pm \ell^\mp$}. Pre-selection: four leptons with total charge $Q=\pm 2$, two of them with $p_T > 30$ GeV. Selection and reconstruction: reconstructed $E$ mass $m_E^\text{rec}$ between 280 and 320 GeV.
\item[(d)] {\em Four leptons $\ell^+ \ell^+ \ell^- \ell^-$}. Pre-selection: four leptons with total charge $Q=0$, two of them with $p_T > 30$ GeV. Selection: events do not have two opposite sign pairs both with an invariant mass closer to $M_Z$ than 5 GeV. Reconstruction: $m_E^\text{rec}$ between 280 and 320 GeV.
\item[(e)] {\em Like-sign trileptons $\ell^\pm \ell^\pm \ell^\pm$}. Pre-selection: three like-sign leptons, two of them with $p_T > 30$ GeV. Selection: $p_T > 50$ GeV for the leading and sub-leading leptons.
\item[(f)] {\em Three leptons $\ell^\pm \ell^\pm \ell^\mp$}. Pre-selection: three leptons with total charge $Q=\pm 1$, the two like-sign ones with $p_T > 30$ GeV. Selection: none of the two opposite-sign pairs has invariant mass closer to $M_Z$ than 10 GeV. Reconstruction: two additional jets with $p_T > 20$ GeV; $m_E^\text{rec}$ and $m_N^\text{rec}$ between 240 and 360 GeV.
\item[(g)] {\em Like-sign dileptons $\ell^\pm \ell^\pm$}. Pre-selection: two like-sign leptons with $p_T > 30$ GeV. Selection: $\ptmiss < 30$ GeV; four jets with $p_T > 20$ GeV. Reconstruction: both heavy lepton invariant masses $m_{\ell_1jj}$, $m_{\ell_2 jj}$ between 250 and 350 GeV.
\item[(h)] {\em Opposite-sign leptons $\ell^+ \ell^-$}. Pre-selection: two opposite-sign leptons with $p_T > 30$ GeV; four jets with $p_T > 20$ GeV. Selection: $\ptmiss < 30$ GeV; leading lepton momentum $p_T^{\ell_1} > 100$ GeV; the two-jet invariant masses $m_{j_1 j_2}$ and $m_{j_3 j_4}$ which reconstruct the two bosons must be between 50 and 150 GeV. Reconstruction: both heavy lepton invariant masses $m_{\ell_1jj}$, $m_{\ell_2 jj}$ between 260 and 340 GeV.
\item[(i)] {\em One charged lepton $\ell^\pm$}. Pre-selection: one charged lepton with $p_T> 30$ GeV; $\ptmiss > 50$ GeV; the charged lepton and missing energy must have transverse mass larger than 200 GeV; four jets with $p_T > 20$ GeV, and no $b$-tagged jets. Selection: charged lepton with $p_T^{\ell} > 100$ GeV; the two-jet invariant masses $m_{j_1 j_2}$ and $m_{j_3 j_4}$ which reconstruct the two bosons must be between 50 and 150 GeV. Reconstruction: $m_{\ell jj}$ between 260 and 340 GeV; $m_{\nu jj}$ between 200 and 400 GeV.
\end{itemize}
The details of the mass reconstructions and kinematical distributions are very lengthy to be reproduced here, but can be found in Ref.~\cite{delAguila:2008cj}.

The number of events with 30 fb$^{-1}$ after these criteria is collected in Table~\ref{tab:comp-T} for the fermion triplet signals in case that the heavy neutrino has Majorana or Dirac character. (Note that for a Dirac triplet the total production cross sections are twice larger.)
The most relevant differences with respect to the Majorana case are the enhancement (by more than a factor of three) of the signal in the $\ell^+ \ell^+ \ell^- \ell^-$
final state and the absence of a significant like-sign dilepton signal.
\begin{table}[t]
\begin{center}
\begin{small}
\begin{tabular}{cccccccccc}
           & $6 \ell$
           & $5 \ell$ & $\ell^\pm \ell^\pm \ell^\pm \ell^\mp$
           & $\ell^+ \ell^+ \ell^- \ell^-$ & $\ell^\pm \ell^\pm \ell^\pm$
           & $\ell^\pm \ell^\pm \ell^\mp$ & $\ell^\pm \ell^\pm$
           & $\ell^+ \ell^-$ & $\ell^\pm$ \\
\hline
$\Sigma$ (M) & 0.6 &  10.6  & 17.4 & 55.7  & 10.2 & 110.3 & 177.8 & 178.7 & 232.4 \\
$\Sigma$ (D) & 1.9 &  21.4  & 9.1  & 173.4 & 2.9  & 194.4 & 4.4   & 607.0 & 314.9 \\
SM Bkg       & 0.0 &  0.9   & 2.5  & 14.3  & 1.9  & 15.9  & 19.5  & 548.3 & 1328
\end{tabular}
\end{small}
\caption{Number of events with 30 fb$^{-1}$ for the fermion triplet signals with Majorana (M) and Dirac (D) neutrinos, and SM background in different final states.}
\label{tab:comp-T}
\end{center}
\end{table}
In the
$\ell^\pm \ell^\pm \ell^\mp$ mode the Dirac signal is 1.75 times larger, too, which makes this channel the most interesting for Dirac neutrino triplet discovery. Then, a fermion triplet with a Majorana neutrino is characterised by the presence of trilepton and like-sign dilepton signals with similar significance, and a smaller $\ell^+ \ell^+ \ell^- \ell^-$ signal~\cite{delAguila:2008cj} while if the heavy neutrino has Dirac nature the like-sign dilepton signal is absent and the four lepton signal has the same significance as the trilepton one. For the triplet mass assumed, $5\sigma$ significance with a heavy Dirac neutrino can be achieved with 1.5 and 1.7 fb$^{-1}$
in the $\ell^\pm \ell^\pm \ell^\mp$ and $\ell^+ \ell^+ \ell^- \ell^-$ final states, respectively. Further interesting differences are found in some of the remaining channels. In the five lepton channel, the Dirac signal could be observed with 14 fb$^{-1}$, half the luminosity required for the Majorana case. The opposite-sign lepton channel is also enhanced in the Dirac case (as expected) but the background is larger and it cannot compete with the cleanest modes ($5\sigma$ for 6 fb$^{-1}$).
Conversely, the $\ell^\pm \ell^\pm \ell^\pm \ell^\mp$
and $\ell^\pm \ell^\pm \ell^\pm$ channels, for which the main contributions to the signals involve LNV neutrino decays, are suppressed if the heavy neutrino has Dirac character.

\section{Summary}

In this Letter we consider the Dirac variant of seesaw type III, in which one or more heavy neutrinos are Dirac particles. We write down the interactions of the new heavy Dirac neutrino $N$ and its two charged lepton partners $E_1^-$, $E_2^+$. It has also been shown that, analogously to seesaw I, heavy quasi-Dirac neutrinos arise naturally if one assigns lepton number $\text{L}=1$, $\text{L}=-1$ to two fermion triplets and requires that lepton number breaking is small in the heavy sector.
We have then studied the observability of heavy (quasi-)Dirac neutrino singlets and triplets.

Our analysis shows that Dirac and Majorana neutrino singlets and triplets can be clearly distinguished by looking for their signals in different final states. The differences are apparent, as it can be observed in Tables~\ref{tab:comp-S} and \ref{tab:comp-T}. We have to remember that the total cross section for a Dirac triplet (composed by two Majorana triplets) is two times larger than for a Majorana one. In the singlet case the charged current couplings are assumed to be the same for heavy Dirac and Majorana neutrinos, and equal to the upper limit from precision electroweak data, and then the production cross sections are the same too.
In some final states the signal enhancements for a Dirac neutrino (compared to a Majorana one) can be up to a factor of three, while some final states present in the Majorana case, namely like-sign dileptons, are absent or very small. The kinematics and distributions of the new signals are very similar to the corresponding ones for Majorana neutrinos, presented with great detail in Ref.~\cite{delAguila:2008cj}, with the exception of the LNV final states which are not produced in the Dirac case.

We have found the somewhat unexpected result that heavy Dirac neutrino singlets coupling to the muon could be observable at LHC in the trilepton channel $\ell^\pm \ell^\pm \ell^\mp$, achieving $5\sigma$ significance with a luminosity of only 13 fb$^{-1}$ for $m_N = 100$ GeV, $V_{\mu N}^2
= 0.0032$. For heavy triplets the prospects are much better, with discovery
for only 1.5 fb$^{-1}$ for a triplet mass of 300 GeV in the trilepton channel alone.
This excellent discovery potential confirms the trilepton channel as the most interesting for the investigation of seesaw III at LHC, in the Dirac as well as in the Majorana case (there the sensitivity is slightly worse than for the like-sign dilepton final state but the presence of a heavy neutral lepton can be established with the mass reconstruction and charge determination). 

It is worth remarking that the heavy Dirac neutrino discovery potential depends on very few parameters. In the case of neutrino singlets these are the heavy neutrino mass and its mixing angle. While the dependence on the mixing angle is trivial because cross sections scale with $|V_{lN}|^2$, the dependence on the neutrino mass is more complicated, resulting from the interplay of several factors: (i) the total production cross section, which decreases with the mass; (ii) decay branching ratios, where for $m_N > M_H$ a new decay channel opens reducing the like-sign dilepton and trilepton signals; (iii) detection efficiencies, which significantly grow with $m_N$ in the range studied; (iv) the SM background, which is relatively large
in the region of small transverse momenta, and whose size also depends on the kinematics of the signal. Hence, although an accurate estimate requires a detailed simulation, an educated guess gives a discovery reach slightly above $m_N = 100$ GeV with 30 fb$^{-1}$.

In the triplet case only the mass is relevant because the production cross section does not depend on the mixing (it only determines possible vertex displacements if it is very small~\cite{Franceschini:2008pz}). Moreover, in this case the SM background is tiny and the discovery potential is essentially determined by the production cross section, whose dependence on the common triplet mass $m_\Sigma$ is shown in Fig.~\ref{fig:cross} (right). Thus, masses up to approximately 700 GeV could be discovered with 30 fb$^{-1}$. This discovery limit is slightly lower than the one estimated for a Majorana triplet (750 GeV) despite the production cross section being twice larger. The reason for this apparent paradox is simple: for a Dirac triplet the LNC signatures are enhanced with respect to the LNV ones, as it can be observed in Table~\ref{tab:comp-T}. Then, the signal in the cleanest channels, which determine the discovery potential, is almost the same in both cases for equal values of the mass.
Finally, it should be stressed that the main objective of the simulations presented here is to show how the Dirac and Majorana cases could be distinguished if a positive signal is observed. In this way, this study complements the general analysis in Ref.~\cite{delAguila:2008cj} for minimal seesaw models with Majorana neutrinos.

\vspace{1cm}
%\newpage
\noindent
{\Large \bf Acknowledgements}
\vspace{0.3cm}

\noindent
This work has been supported by MEC project FPA2006-05294 and
Junta de Andaluc{\'\i}a projects FQM 101, FQM 437 and FQM03048. The work of J.A.A.S.
has been supported by a MEC Ram\'on y Cajal contract.

\newpage
\appendix

\section{Feynman rules}

We give here the Feynman rules used in our matrix element calculations. Lepton number is conserved except for effects proportional to light neutrino masses (which can be taken to be zero at high energy), and all vertices conserve lepton number, with the peculiarity that for the charged lepton $E_2^+$ the particle is positively charged and the antiparticle has negative charge.

\begin{table}[htb]
\begin{center}
\begin{footnotesize}
\begin{tabular}{clcclcl}
\raisebox{-11mm}{\epsfig{file=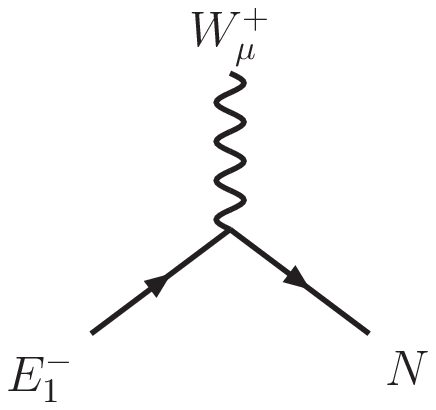,height=22mm,clip=}}
  & $\displaystyle -ig \gamma^\mu$ & \quad &
\raisebox{-11mm}{\epsfig{file=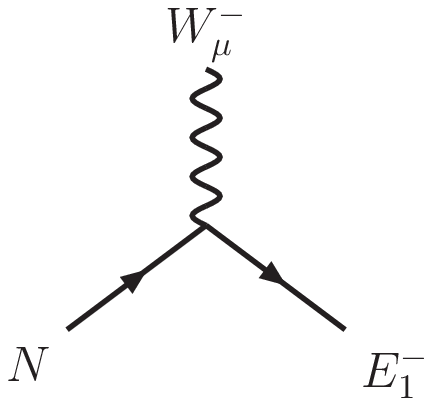,height=22mm,clip=}}
  & $\displaystyle -ig \gamma^\mu$ \\ \\
\raisebox{-11mm}{\epsfig{file=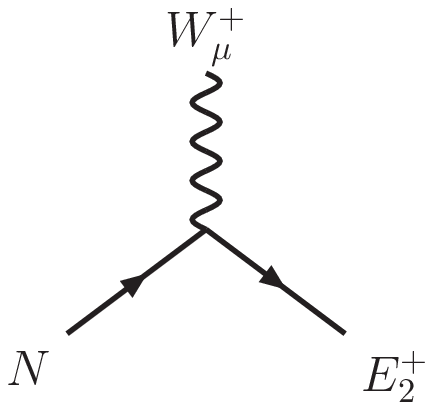,height=22mm,clip=}}
  & $\displaystyle ig \gamma^\mu$ & \quad &
\raisebox{-11mm}{\epsfig{file=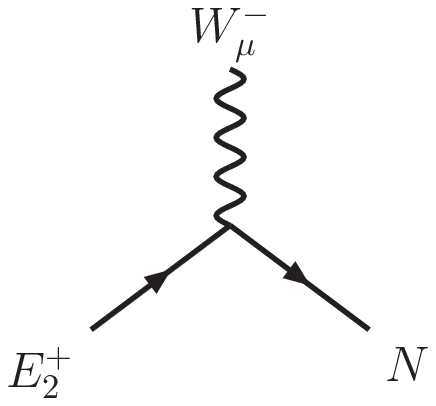,height=22mm,clip=}}
  & $\displaystyle ig \gamma^\mu$ \\ \\
\raisebox{-11mm}{\epsfig{file=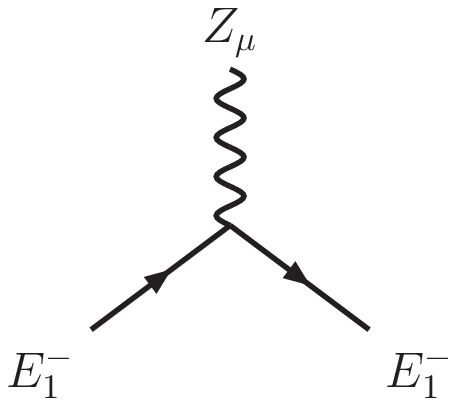,height=22mm,clip=}}
  & $\displaystyle igc_W \gamma^\mu$ & \quad &
\raisebox{-11mm}{\epsfig{file=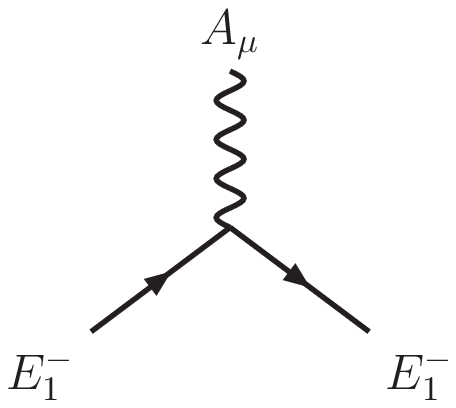,height=22mm,clip=}}
  & $\displaystyle ie \gamma^\mu$ \\ \\
\raisebox{-11mm}{\epsfig{file=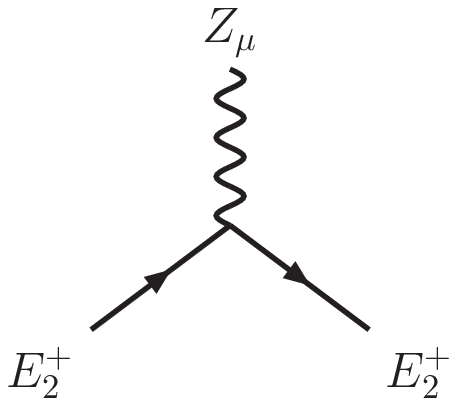,height=22mm,clip=}}
  & $\displaystyle -igc_W \gamma^\mu$ & \quad &
\raisebox{-11mm}{\epsfig{file=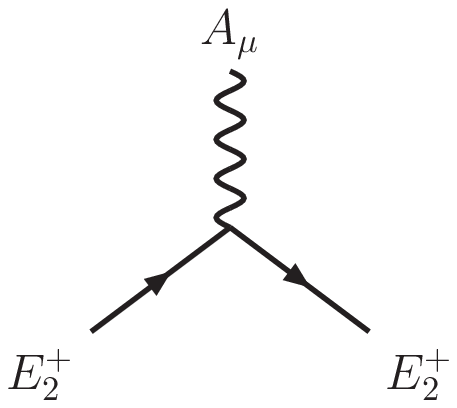,height=22mm,clip=}}
  & $\displaystyle -ie \gamma^\mu$ \\ \\
\end{tabular}
\end{footnotesize}
\caption{Feynman rules for heavy fermion triplet gauge interactions. For clarity, in charged current interactions we indicate the charge of the incoming $W$ boson.}
\end{center}
\end{table}

\begin{table}[p]
\begin{center}
\begin{footnotesize}
\begin{tabular}{clccl}
\raisebox{-11mm}{\epsfig{file=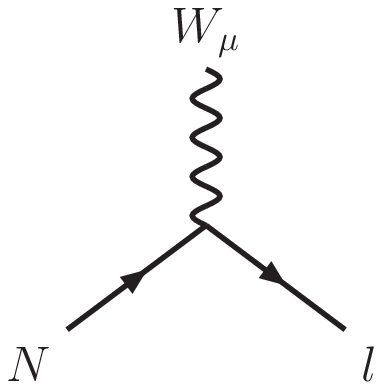,height=22mm,clip=}}
  & $\displaystyle -\frac{ig}{\sqrt 2} V_{lN} \gamma^\mu P_L$ & \quad &
\raisebox{-11mm}{\epsfig{file=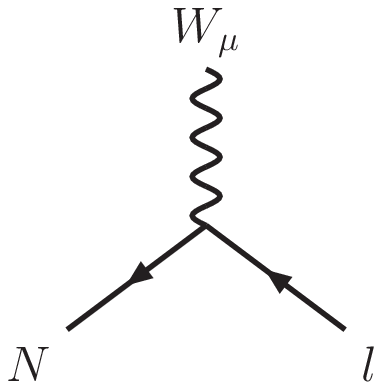,height=22mm,clip=}}
  & $\displaystyle -\frac{ig}{\sqrt 2} V_{lN}^* \gamma^\mu P_L$ \\ \\
\raisebox{-11mm}{\epsfig{file=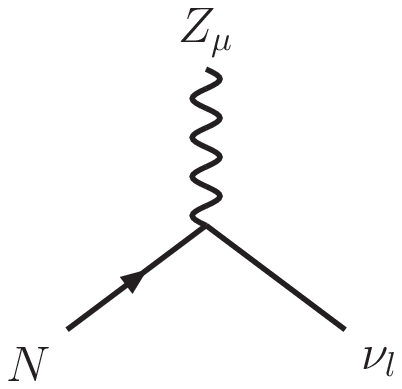,height=22mm,clip=}}
  & $\displaystyle - \eta \frac{ig}{2c_W} V_{lN} \gamma^\mu
   P_L $ & \quad &
\raisebox{-11mm}{\epsfig{file=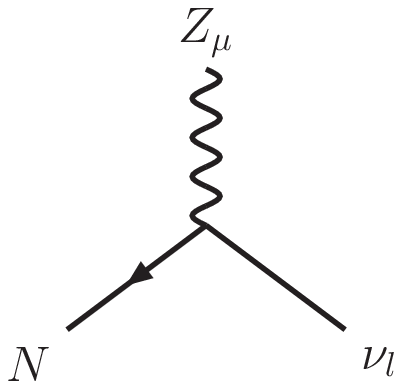,height=22mm,clip=}}
  & $\displaystyle - \eta \frac{ig}{2c_W} V_{lN}^*  \gamma^\mu 
  P_L $ \\ \\
\raisebox{-11mm}{\epsfig{file=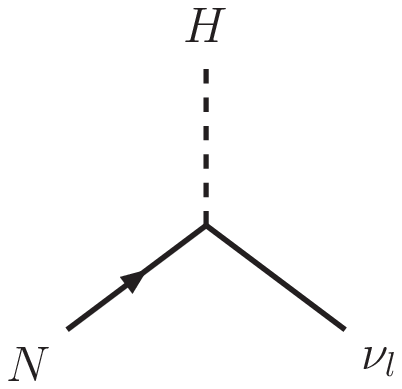,height=22mm,clip=}}
  & $\displaystyle - \eta \frac{ig m_N}{2 M_W} V_{lN} P_R$ & \quad &
\raisebox{-11mm}{\epsfig{file=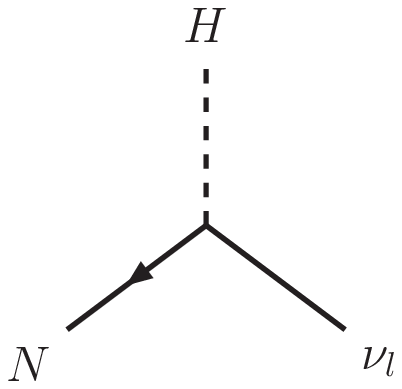,height=22mm,clip=}}
  & $\displaystyle - \eta \frac{ig m_N}{2 M_W} V_{lN}^* P_L $ \\ \\
\end{tabular}
\end{footnotesize}
\caption{Feynman rules for heavy Dirac neutrino singlet ($\eta=1$) and triplet
($\eta=-1$) interactions with SM fermions.}
\end{center}
\end{table}

\begin{table}[p]
\begin{center}
\begin{footnotesize}
\begin{tabular}{clccl}
\raisebox{-11mm}{\epsfig{file=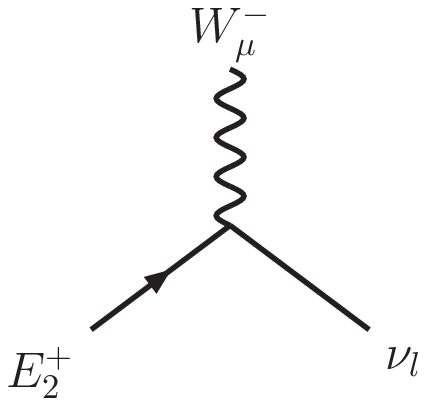,height=22mm,clip=}}
  & $\displaystyle ig V_{lN} \gamma^\mu P_L$ & \quad &
\raisebox{-11mm}{\epsfig{file=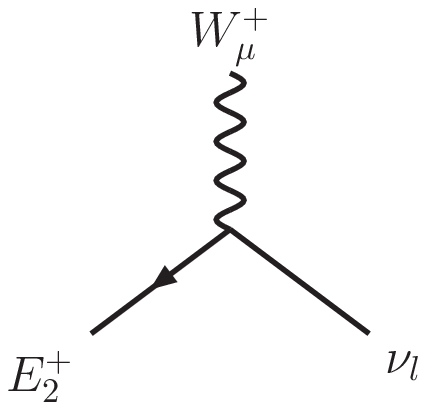,height=22mm,clip=}}
& $\displaystyle ig V_{lN}^* \gamma^\mu P_L$ \\ \\
\raisebox{-11mm}{\epsfig{file=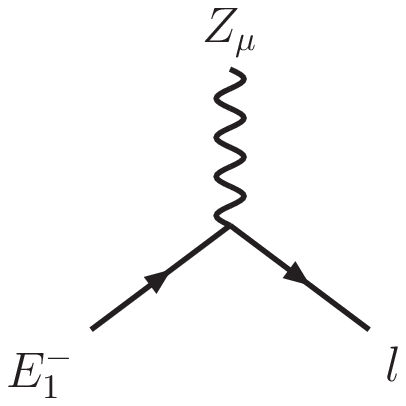,height=22mm,clip=}}
  & $\displaystyle \frac{ig}{\sqrt 2c_W} V_{lN} \gamma^\mu  P_L  $ & \quad &
\raisebox{-11mm}{\epsfig{file=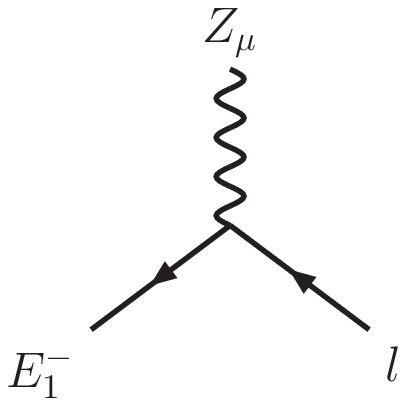,height=22mm,clip=}}
  & $\displaystyle \frac{ig}{\sqrt 2c_W} V_{lN}^* \gamma^\mu P_L  $ \\ \\
\raisebox{-11mm}{\epsfig{file=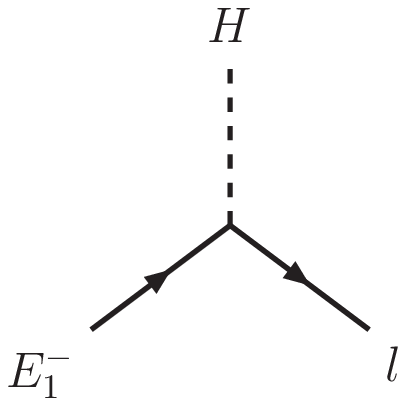,height=22mm,clip=}}
  & $\displaystyle \frac{ig m_E}{\sqrt 2 M_W} V_{lN} P_R$ & \quad &
\raisebox{-11mm}{\epsfig{file=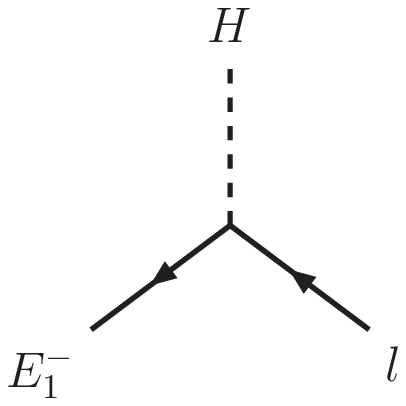,height=22mm,clip=}}
  & $\displaystyle \frac{ig m_E}{\sqrt 2 M_W} V_{lN}^* P_L$ \\ \\
\end{tabular}
\end{footnotesize}
\caption{Feynman rules for heavy charged lepton interactions with SM fermions.
For clarity, in charged current interactions we indicate the charge of the incoming $W$ boson.}
\end{center}
\end{table}

\end{document}